\documentstyle[preprint,aps]{revtex}
\def\journal#1, #2, #3#4#5#6, #7#8    {
    {\em #1~}{\bf #2}, #7#8 (#3#4#5#6)}

\def\prb{\journal Phys. Rev. B, }

\def\prl{\journal Phys. Rev. Lett., }

\def\Bb{{I\!\! B}}

\newcommand{\beq}[1]{\begin{equation}\label{#1}}
\newcommand\eeq{\end{equation}}
\newcommand{\ba}[1]{\begin{eqnarray}\label{#1}}
\newcommand{\baa}{\begin{eqnarray}}
\newcommand\ea{\end{eqnarray}}
\newcommand{\bee}{\begin{equation}}
\def\nn{\nonumber \\}
\def\l{\lambda}
\def\a{\alpha}
\def\f{\varphi}

\newcommand{\h}{Hamiltonian}
\newcommand{\an}{angular momentum}
\newcommand{\B}[1]{{\bf #1}}
\def\hlf{\frac{1}{2}}

\newcommand{\lsim}{\stackrel{\rm <} {\scriptstyle \sim}}

\begin{document}
\title{Analytical Results for 
 Trapped Weakly Interacting Bosons in Two Dimensions}
\author{Velimir Bardek and Stjepan Meljanac
\footnote{e-mail address: 
bardek@thphys.irb.hr \\ \hspace*{3cm} meljanac@thphys.irb.hr} }
\address{Theoretical Physics Division,\\
Rudjer Bo\v skovi\'c Institute, P.O. Box 180,\\
HR-10002 Zagreb, CROATIA}
\maketitle
\begin{abstract}
We consider a model of $N$ two-dimensional bosons in a harmonic trap with 
translational and rotational invariant, weak two-particle interaction. We
present in configuration space 
a systematical recursive method for constructing all wave functions 
with angular momentum $L$ and corresponding energies and apply it 
to $L\leq 6 $ for all 
$N$. The lower and the upper bounds for interaction energy 
are estimated. We analitically confirm 
the conjecture 
of Smith et al. that elementary symmetric polynomial is the ground state 
for repulsive delta interaction,
for all $N\geq L$ up to $L\leq 6 $. Additionally, we find that 
there exist vanishing-energy solutions for $L\geq N(N-1)$, 
signalizing the exclusive statistics. Finally, we consider 
briefly the case of attractive power-like  potential $r^k,\;k>-2$, and 
prove that the lowest-energy state is still the one in which all 
\an \ is absorbed by the center-of-mass motion.
\end{abstract}
\vspace{1cm}

\newpage

Considerable attention has been devoted to the fenomenon of Bose-Einstein (BE)
condensation in atomic vapors in the past few years\cite{1,2,3}. 
The main reason was the 
fact that such systems might form quantized vortices under rotation, like in 
superfluid $^4$He.  
Many authors considered the vortex problem 
in BE  condensates theoretically\cite{4,5,6,7,8,9}. In the meantime, 
the exsitence of vortex state has been
experimentally confirmed\cite{10,11}.

Persuing an analogy with the fractional quantum Hall (FQH) effect\cite{12}
Wilkin, Gunn, and Smith introduced a model of weakly interacting bosons in a
harmonic trap\cite{5}. 
In the absence of interactions, there is a huge degeneracy 
corresponding to the number of ways to distribute $L$ units of \an \ among 
$N$ bosons. The degeneracy can be lifted by diagonalizing the interaction 
in the restricted Hilbert space of the lowest-energy states with \an \ $L$.
Accordingly, Bertsch and Papenbrock used numerical computation to 
show that the ground-state energy depends linearly on \an \ $L$ and that 
the corresponding ground state of the repulsive model for $L\leq N$ is 
the elementary symmetric polinomial of coordinates $z_i=x_i+iy_i$, relative 
to the center-of-mass, $R=\frac{1}{N}\sum_iz_i$\cite{7}.  

Recently, Smith and Wilkin proved analytically that these symmetric polynomials 
are indeed exact eigenstates\cite{13}. Later, Huang\cite{14}
 extended the work by 
Smith and Wilkin to the models interacting via arbitrary rotationally and 
translationally symmetric potentials. The main result was that the associated 
interaction energy  still varies linearly with $L$. In addition, a lower 
bound for the attractive quartic 
interaction energy was derived. Finally, Bertsch and Papenbrock 
showed\cite{15} that there is a subspace structure that explains  why certain 
eigenstates turn out to be simple analytical functions.
They succeeded in constructing basis states but only up to $L=5$.
The mentioned subspace structure is present in a rather wide class of 
two-body interactions.
However, for general interaction, the problem of determining the ground state 
has  not yet been solved analytically.

Our starting point is the two-dimensional \h \ given by
\beq 1
H=\sum_{i=1}^N\left(-\hlf{\B\nabla_i}^2+\hlf\B r_i^2\right)+\sum_{i<j}^N
v(|\B r_i-\B r_j|).\eeq
It describes $N$ Bose-Einstein particles in a harmonic trap, weakly interacting 
via the two-particle potential $v$, possesing translational and rotational 
symmetries. Since the spacing between noninteracting energy levels is 
greater then the two-body interaction strength, we restrict the Hilbert 
space to the degenerate ground-state manifold of $N$ non-interacting bosons.
 The single particle state is of the form $z^m\exp(-\hlf|z|^2)$ where 
$z=x+iy$ and $m$ is the \an \ quantum number. The eigenfunctions of the 
many-body problem are
\beq 2
\psi(z_1,z_2,\ldots,z_N)=\f(z_1,z_2,\ldots,z_N)\prod_{i=1}^N
\exp(-\hlf|z_i|^2), \eeq
where $\f$ is a homogeneous polynomial of degree $L$ that is totally symmetric 
under permutation of particle indices. Suitable basis functions for such 
polynomials are given by
\beq 3
B_{\l}(z_1,z_2,\ldots,z_N)=\frac{1}{n_1!n_2!\cdots n_p!}
\sum_{i_1,i_2,\ldots,i_q=1}^N\hspace{-0.6cm}^{'}
z_{i_1}^{\l_1}z_{i_2}^{\l_2}\cdots z_{i_q}^{\l_q},\eeq
where set $\{\l_1,\l_2,\ldots,\l_q\}$ denotes any partition of $L$ such that 
$\sum_{i=1}^q\l_i=L$ and $\l_1\geq\l_2\geq\cdots\geq\l_q>0$ for 
$q\leq N$\cite{16}. 
The prime denotes the summ over mutually different indices 
$i_1,i_2,\ldots,i_q$, 
while the numbers $n_1,n_2,\ldots,n_p$ denote the frequencies of appearences 
of equal $\l's$ (i. e., the number of particles carrying the same \an).
Note that the number of distinct monomial terms 
$z_{i_1}^{\l_1}z_{i_2}^{\l_2}\cdots z_{i_q}^{\l_q}$ in $B_{\l}$ is given by
$N(N-1)\cdots(N-q+1)/(n_1!n_2!\cdots n_p!)$, where $n_1+n_2+\cdots+n_p=q$.
For simplicity we omit the exponentials from wavefunctions throughtout the 
paper.
Owing to the translational and rotational symmetries of the two-particle 
interaction $v$, we find, for  non-negative integers $n$ and $m$
\beq 4
v(z_1+z_2)^n(z_1-z_2)^{2m}P(z_3,z_4,\ldots,z_N)=c_{2m}(z_1+z_2)^n(z_1-z_2)^{2m}
P(z_3,z_4,\ldots,z_N), \eeq
where $P$ denotes an arbitrary polynomial in $z_3,z_4,\ldots,z_N$ variables, 
and $c_n$ is given by
\beq 5
c_n=\frac{\int_0^{\infty}dr\, r^{2n+1}v(r)\exp(-r^2/2)}
{\int_0^{\infty}dr\, r^{2n+1}\exp(-r^2/2)}\; .\eeq 
It is evident that the coefficients $c_n$ represent the interaction energy 
$v(r)$ of the relative motion of two bosons in the single particle state 
$r^n\exp(-r^2/2)$ with \an \ $n$.
We are now in position to estimate the lower and the upper bound of the 
interaction $V=\sum_{i<j}^Nv(|\B r_i-\B r_j|)$. Since $c_{\rm min}\leq\bar v
\leq c_{\rm max}$ where $c_{\rm min(max)}={\rm min(max)}\{c_0,c_2,\ldots,c_{2
\left[\frac{L}{2}\right]}\},$ the total interaction energy lies between
\beq 6
{N\choose 2}c_{\rm min}\leq\overline{V}
\leq {N\choose 2} c_{\rm max} .\eeq
In order to simplify the calculations concerning the action of the potential 
$V$ on polynomials $B_{\l}$, let us define symmetric functions of two variables:
\beq 7
b_{ij}(z_1,z_2)=\hlf(z_1^iz_2^j+z_1^jz_2^i),\;i\geq j.\eeq
The action of the potential $v$ on $b_{ij}$ is given by:
\beq 8
v\,b_{ij}=\sum_{l=0}^{[\frac{n}{2}]}\alpha_{ij}^{kl}b_{kl},\eeq
where $i+j=k+l=n$, and the coefficients $\alpha_{ij}^{kl}$ satisfy the 
summation rule $\sum_{l=0}^{[\frac{n}{2}]}\alpha_{ij}^{kl}=c_0$.
The first constraint correspondes to the conservation of total \an \ for a
rotationally symmetric potential, while the summation rule reflects the 
presence of translational symmetry.
By using the simple calculations it can be shown that 
\beq 9
\alpha_{ij}^{kl}=\frac{2-\delta_{kl}}{2^n}\sum_{p=0}^{[\frac{n}{2}]}
c_{2p}S_{i,j}^{2p}S_{n-2p,2p}^l, \eeq
where
\beq a
S_{i,j}^q=\sum_{r+s=q}(-)^s{i\choose r}
{j\choose s} .\eeq
The coefficients $\alpha_{ij}^{kl}$ represent the two-body matrix element 
$V_{ijkl}$ (see Ref.\cite{13})
of the interaction potential $V$. The relation (\ref{9}) expresses the 
coefficients $\alpha_{ij}^{kl}$ in terms of interaction energies $c_n$
of a pair of bosons having the \an \ $n$.
It can be verified that in the case of constant potential $v=v_{\rm const}=
c_{2n}$ 
and $\alpha_{ij}^{kl}$ reduces to $\alpha_{ij}^{kl}=\delta_i^k\delta_j^l$.
Applying potential $V$ onto $B_1^n(z_1,z_2,\ldots,z_N)$ and using the 
relation (\ref{4}), we obtain
\beq b 
V\,B_1^n(z_1,z_2,\ldots,z_N)=c_0{N\choose 2}B_1^n(z_1,z_2,\ldots,z_N), \eeq
i. e., the polynomial $B_1^n(z_1,z_2,\ldots,z_N)$ is an eigenvector with 
eigenvalue equal to the lowest (greatest) interaction energy, if 
$c_0=c_{\rm min(max)}$. Furthermore, the action of the potential $V$ on 
the product $B_1^nB_{\l}$ reduces to
\beq c
V\,B_1^nB_{\l}=B_1^nVB_{\l}, \eeq
for any $n$ and partition $\l$.
Owing to the property (\ref{c}) it is convinient to choose the natural basis 
for a given total \an \ $L$ as a set formed by $B_1^L$ and products
$B_1^{L-K}B_{\l}$ where $2\leq K\leq L$, and $\l$ stands for the special 
partition of $K$; $\l=\{\underbrace{1,1,\ldots,1}_{K\;{\rm times}}\}\equiv 1^K$
and all other partitions not containing 1's i. e., 
$\l=\{\l_1,\l_2,\ldots,\l_k\},\;\l_1\geq\l_2\geq\cdots\l_k\geq 2$, 
where $\sum_{i=1}^k\l_i=K$ and $2\leq k\leq N$.
For $K>N$ the partition $\l=1^K$ is not present and for $K=N+1,N+2$ the 
remaining set of polynomials $B_{\l}$ is linearly independent. For 
$K>N+2$ the above states become linearly dependent\cite{16}.
Let us call the corresponding subspace spanned by above defined $B_{\l}$ 
polynomials as $I\!\!B_K$. For example $I\!\!B_2={\rm span}\{B_{1^2}\}$, 
$I\!\!B_3={\rm span}\{B_{1^3}\}$, $I\!\!B_4={\rm span}\{B_{1^4},B_{22}\}$,
 $I\!\!B_5={\rm span}\{B_{1^5},B_{32}\}$, $I\!\!B_6={\rm span}\{B_{1^6},B_{42},
B_{33},B_{222}\}$, etc.
The whole $L=6$ space can be described as a summ of appropriate subspaces:
$B_1^6\bigoplus B_1^4I\!\!B_2\bigoplus B_1^3I\!\!B_3\bigoplus
B_1^2I\!\!B_4\bigoplus B_1I\!\!B_5\bigoplus I\!\!B_6$.
We note in passing that these subspaces are not mutually orthogonal.

Let us now turn to the solution of the eigenvalue problem. 
Having in mind the translational and rotational invariance of the potential 
$V$ it is obviuos that $VB_1^{L-K}\Bb_K=B_1^{L-K}V\Bb_K$ and 
$V\Bb_K\subset {\rm span}\{V\Bb_K,B_1V\Bb_{K-1},B_1^2V\Bb_{K-2},\ldots,B_1^K\}$.
These give rise to the following sequence of the subspace relations:
\ba d
&&VB_1^L=c_0{N\choose 2}B_1^L, \nn
&&VB_1^{L-2}\Bb_2\subset {\rm span}\{B_1^L,B_1^{L-2}\Bb_2\},\nn
&&VB_1^{L-3}\Bb_3\subset {\rm span}\{B_1^L,B_1^{L-2}\Bb_2,B_1^{L-3}\Bb_3\},\nn
&&\hspace{2cm}\vdots \ea
Consequently, the matrix of potential $V$ in the natural basis possesses the 
block triangular form which means that the original eigenvalue problem 
has been substantionally simplified, i. e., reduced to the eigenvalue 
problem in the $\Bb_K$ subspaces, for $2\leq K\leq L$.
Hence, we have to calculate $VB_{1^L},\,VB_{22},\,VB_{32},\,VB_{42},\,VB_{33},\,
VB_{222}$, etc., then express the result in terms of our natural basis and 
finally project them onto the $\Bb_K$ subspace.
Applying the potential $V$ onto the state $B_{1^L}$ yields:
\ba e
&&VB_{1^L}(z_1,z_2,\ldots,z_N)=V\sum_{i<j}^N\left[z_iz_j{N-2\choose L-2}_t
+(z_i+z_j){N-2\choose L-1}_t+{N-2\choose L}_t\right] \nn
&&= \left[ \alpha_{11}^{11}\frac{L(L-1)}{2}+c_0L(N-L)+\frac{c_0}{2}
(N-L)(N-L-1)\right] B_{1^L}+\alpha_{11}^{20}\frac{N-L+1}{2}B_{21^{L-2}},\ea
where the coefficients $\alpha$ are given by relation (\ref{9}):
\beq f
\alpha_{11}^{11}=\frac{c_0+c_2}{2},\;\alpha_{11}^{20}=\frac{c_0-c_2}{2}.\eeq
In (\ref{e}) we have explicitly extracted the $(z_i,z_j)$ pair of 
coordinates to simplify the action of the potential $V$.
Binomial coefficients with index $t$ symbolically denote the number of 
terms in remaining $N-2$ variables. By using the identity:
\beq g 
B_{21^{L-2}}=B_1B_{1^{L-1}}-LB_{1^L},\eeq
we obtain 
\beq h
VB_{1^L}(z_1,z_2,\ldots,z_N)=\frac{c_0-c_2}{4}(N-L+1)B_1B_{1^{L-1}}+\left[
c_0{N\choose 2}-\frac{c_0-c_2}{4}NL\right]B_{1^L}.\eeq
Hence, the eigenvalue is given by:
\beq j
\Lambda_{1^L}=c_0{N\choose 2}-\frac{c_0-c_2}{4}NL.\eeq
Having find the eigenvalue we are now in the position to find the 
corresponding eigenfunction. By expanding the unknown eigenvector in terms of 
basis polynomials $B_{1^L}$ and $B_1^{L-k}B_k$ with unknown coefficients, we 
easily obtain the system of simple recursive relations. Solving these 
equations we finally get the expansion coefficients in which all dependence 
on the details of interaction $V$, namely on $c_0$ and $c_2$, simply drops out!
 The corresponding eigenvector is:
\beq k
A_{1^L}=\sum_{n=0}^L(-)^nB_{1^{L-n}}\left(\frac{B_1}{N}\right)^n=
B_{1^L}(z_1-\frac{B_1}{N},z_2-\frac{B_1}{N},\ldots,z_N-\frac{B_1}{N}),\eeq
in agreement with Ref.\cite{17}.
In the same way we find:
\beq l
VB_{22}=\left[\a_{22}^{22}+2\a_{20}^{20}(N-2)+c_0{N-2\choose 2}\right]B_{22}
+\hlf\a_{22}^{31}B_{31}+\frac{N-1}{2}\a_{22}^{40}B_{4}+2\a_{20}^{11}B_{211}
,\eeq
\ba m
VB_{32}&=&\left[\a_{32}^{32}+\a_{30}^{30}(N-2)+\a_{20}^{20}(N-2)+
c_0{N-2\choose 2}\right]B_{32}\nn
&+&\a_{32}^{41}B_{41}+(N-1)\a_{32}^{50}B_{5}+2\a_{30}^{21}B_{221}+
2\a_{20}^{11}B_{311},\ea
\ba n
VB_{42}&=&\left[\a_{42}^{42}+\a_{40}^{40}(N-2)+\a_{20}^{20}(N-2)+
c_0{N-2\choose 2}\right]B_{42}\nn
&+&2\a_{42}^{33}B_{33}
+\a_{42}^{51}B_{51}+(N-1)\a_{42}^{60}B_{6}+\a_{40}^{31}B_{321}+
6\a_{40}^{22}B_{222}+2\a_{20}^{11}B_{411},\ea
\ba o
VB_{33}&=&\left[\a_{33}^{33}+2\a_{30}^{30}(N-2)+c_0{N-2\choose 2}\right]B_{33}
+\hlf\a_{33}^{42}B_{42}\nn &+&\hlf\a_{33}^{51}B_{51}+
\frac{N-1}{2}\a_{33}^{60}B_{6}+\a_{30}^{21}B_{321},\ea
\beq p
VB_{222}=\left[3\a_{22}^{22}+3\a_{20}^{20}(N-3)+c_0{N-3\choose 2}\right]B_{222}
+\hlf\a_{22}^{31}B_{321}+\frac{N-2}{2}\a_{22}^{40}B_{42}+2\a_{20}^{11}B_{2211}
.\eeq
In order to express the above results in terms of the vectors of our 
natural basis we multiply systematically all polynomials $B_{\l}$ by 
$B_1$ in the following sequence: $B_1^2;\;B_1B_{11},B_1B_2;\;
B_1B_{1^3},B_1B_{21},
B_1B_3;\;B_1B_{1^4},B_1B_{22},B_1B_{211},B_1B_{31},B_1B_4;\,{\rm etc.}$
These products can be easily evaluated by simple algebraic manipulations. 
Here, we give a few results of these manipulations:
\ba q
&&B_1^2=B_2+2B_{11},\nn
&&B_1B_{11}=B_{21}+3B_{1^3},\nn
&&B_1B_2=B_3+B_{21},\nn
&&B_1B_{1^3}=B_{211}+4B_{1^4},\nn
&&B_1B_{211}=B_{31^{2}}+2B_{2^21}+3B_{21^3},\nn
&&\hspace{2cm}\vdots \ea
It is obvious that any polynomial can be expressed in terms of natural basis 
by method of succesive substitution. For example, from the first three
relations we obtain
$B_3=B_1^3-3B_1B_{11}+3B_{1^3}$.

Next, we turn to the final step. Projecting out the states $B_4,B_{31}$ and 
$B_{211}$ onto $B_{22}$, then $B_5,B_{41},B_{31^2}, B_{21^3}$ and 
$B_{221}$ onto $B_{32}$, and $B_6,B_{51},B_{321}, B_{2^21^1}, B_{31^3}, 
B_{21^4}$ and $B_{41^2}$ onto $B_{42}, B_{33}$ and $B_{222}$, we obtain
\ba r
&&P(B_4)=-P(B_{31})=2B_{22},\nn
&&P(B_{211})=0,\nn
&&P(B_5)=-P(B_{41})=5B_{32},\nn
&&P(B_{31^2})=2B_{32},\nn
&&P(B_{21^3})=0,\nn
&&P(B_{221})=-B_{32},\nn
&&P(B_6)=-P(B_{51})=9B_{42}+16B_{33}-18B_{222},\nn
&&P(B_{321})=-B_{42}-2B_{33},\nn
&&P(B_{2^21^1})=-\hlf P(B_{31^3})=\hlf(B_{42}+2B_{33}-3B_{222}),\nn
&&P(B_{21^4})=0,\nn
&&P(B_{41^2})=4B_{42}+8B_{33}-9B_{222},\ea
where $P$ means the projection. Note that vectors $B_{42}, B_{33}$ and 
$B_{222}$ are linearly independent if $N\geq 4$, while for $N=3$ we have:
$P(B_{42})=-2B_{33}+3B_{222}$. 
Upon substitution of results (\ref{r}) into relations (\ref{l}-\ref{p}) 
we find two 
more, yet not known, eigenvalues:
\beq s
\Lambda_{22}=c_0\frac{N-2}{2}(N-\frac{3}{4})+c_2\frac{3N-6}{4}+c_4\frac{N+6}{8},
\;N\geq 2,\;L=4,\eeq
\beq t
\Lambda_{32}=\frac{c_0}{16}\left[17N-36+8{N-2\choose 2}\right]+\frac{c_2}{8}
(5N-12)+\frac{c_4}{16}(5N+2),\;N\geq 3,\;L=5.\eeq
The first one in the case of repulsive $\delta$ potential ($c_0\neq 0$, and 
$c_n=0$, $\forall n>0$) reduces to the result already obtain in Ref.\cite{17}.
The two corresponding eigenstates are obtained in the analogous way as 
$A_{1^L}$:
\beq u
A_{22}=N{\cal B}_4-3{\cal B}_2^2,\eeq
\beq v
A_{32}=N{\cal B}_5+2{\cal B}_3{\cal B}_2,\eeq
where we have used compact notation: ${\cal B}_n=\sum_{i-1}^N(z_i-B_1/N)^n$.
Few comments are in order. The two above eigenvectors have been already found  
in Ref.\cite{15} for a wide class of two-body interactions. They do not depend 
on the details of interaction. But, starting with $L=6$ the wave functions 
and eigenvalues in the subspace spanned by vectors $B_{42}, B_{33}$ and 
$B_{222}$ become dependent  on the details of interaction. 
Generally, for any $L$ it can be seen from relations (\ref{n}-\ref{p}) that the 
action of  the potential $V$ on $\Bb_L$ reduces to 
$VB_{\l}=\sum_{\mu}V_{\l}^{\mu}B_{\mu}$, where the matrix elements have the
structure $V_{\l}^{\mu}=\frac{c_0}{2}\delta_{\l\mu}N^2+\beta_{\l\mu}N+
\gamma_{\l\mu}$. All details of the interaction are incorporated in the 
constants $\beta$ and $\gamma$. For large $N$, the eigenvalues 
of this matrix can be easily found by expansion in $1/N$.
For example, for the weak $\delta$ repulsive potential the corresponding 
$3\times 3$
($N\geq 4,\;L=6$) matrix can be written as
\beq w
c_0\left(\begin{array}{ccc}
\frac{N^2}{2}-\frac{51N}{32}+\frac{15}{4} &\frac{9N}{64}-\frac{3}{2} & 
\frac{N}{16}+\frac{1}{8}\\ \frac{N}{2}+\frac{33}{8} & \frac{N^2}{2}-
\frac{7N}{4}-\frac{15}{16} & \hlf \\ -\frac{9N}{16}-\frac{45}{16} &
-\frac{9N}{32}+\frac{63}{32} & \frac{N^2}{2}-2N+\frac{9}{8} 
\end{array}\right).\eeq
The corresponding eigenenergies can be easily expanded in powers of $1/N$
up to ${\cal O}(1/N)$:
\ba x
\Lambda_6^{(1)}&=&c_0\left(\frac{N^2}{2}-2N+\frac{27}{34}\right), \nn
\Lambda_6^{(2)}&=&c_0\left(\frac{N^2}{2}-\frac{15}{8}N+\frac{30}{13}
\right), \nn
\Lambda_6^{(3)}&=&c_0\left(\frac{N^2}{2}-\frac{47}{32}N+\frac{2955}{3536}
\right). \ea 
For $N\geq 10$ the error is of order $\lsim 1\%$. Having in mind all obtained 
eigenvalues, it turns out that for $L\leq 6$ and arbitrary $N\geq L$, the 
lowest energy is given by $\Lambda_{1^L}$. In this way we
 analytically support 
the conjecture of Smith et. al.\cite{13}, up to $L=6$.
Moreover, the first eigenvalue from the relation (\ref{x}) is in a 
perfect agreement with the numerical result for two excited 
octupole modes\cite{18}.
For $L=N$ it is known that the ground state is a vortex $\psi=\prod_{i=1}^N
(z_i-z_c)$ 1 around $z_c=B_1/N$\cite{5,19}. 
For $L>N$, we find that there exists lower eigenvalue then that given by 
(\ref{j}). For example, for $L=6$ and $N=4$ eigenvalue 
$\Lambda_6^{(1)}=1.38$ is lower then $\Lambda_{1^4}=2$. The same is true for 
$L=6$ and $N=3$, i. e. $\Lambda_6^{(1)}=0$ and $\Lambda_{1^3}=3/4$.
For $N=3$ we find that 
$V(A_{1^2}^pA_{1^3}^q)\sim {\cal B}_{2p+3q}$,
and therefore, for a given $L$ there exists only one eigenstate having 
positive eigenvalue $V{\cal B}_L=[1-(-1/2)^{L-1}]{\cal B}_L$\cite{17}.
From the above examples, it seems that for sufficiently large $L>N$ there 
exists the ground state with energy lower than $\Lambda_{1^N}=c_0N(N-2)/4$. 
Moreover, we find vanishing eigenvalues for $L\geq N(N-1)$. 
As $\delta$ potential gives vanishing $c_n$'s for $n>0$, from 
relation (\ref{4}) the corresponding eigenvectors are given by
\beq z
\psi_0=\prod_{i<j}^N(z_i-z_j)^2\psi_{L-N(N-1)},\eeq
where $\psi_{L-N(N-1)}$ denotes any state in the space of total \an \
equal to $L-N(N-1)$. We note that these states saturate the lower bound in 
inequality (\ref{6}). It is interesting to note that the lowest states (\ref{z})
 are of anyonic (fermionic) type in the sense that  they vanish when 
two particles coincide\cite{20}. 
In other words, particles behave like the hard-core
bosons. This might signalize  some sort of statistical transmutation 
and 0 of exclusion statistics\cite{21}.

We should finally point out that for the attractive, weak two-body potential 
$v(r)=\eta r^k$ (for $k\stackrel{\textstyle >}< 0$ and $\eta
\stackrel{\textstyle >}< 0$, respectively), the coefficient $c_n$ is 
given by
\bee\label{11}
c_n=2^{k/2}\eta\frac{\Gamma(n+1+\frac{k}{2})}{\Gamma(n+1)},\eeq
and is always greater then 
\bee\label{12}
c_0=2^{k/2}\eta\Gamma(1+\frac{k}{2}).\eeq
This means that the state $B_1^L$ is the lowest-energy state for fixed $L$ in
which all \an \ is carried by the center-of-mass motion $z_c$.
For $k=4$ and $c_0=8\eta$ the corresponding energy equals $8{N\choose 2}\eta$
in agreement with the result of Ref.\cite{14}.
The $B_1^L$ remains the lowest-energy eigenstate for the attractive $\delta$ 
potential ($c_0=-\eta/2\pi<c_n=0$) and for all attractive two-particle 
potentials for which $c_0=c_{\rm min}$.
This generalization is a simple consequence of inequality (\ref{6}).
In all of these examples the presence of the uncondensed state $B_1^L$ 
is not a surprise because of the attractive nature of the two-body 
interaction\cite{5}. 

In conclusion, we have developed an analytical recursive method in
configuration space for calculating all eigenstates and corresponding
eigenvalues for trapped bosons interacting via weak, translationally and
rotationally symmetrical
pairwise potentials. We applied the method to states with $L\leq 6$,
 for all $N$.
We have derived 
the lower and the upper bounds for general interaction energy. For
repulsive $\delta-$interaction we have 
proved the conjecture of Smith and Wilkin on
the ground state structure up to $L\leq 6$, for $N\geq L$.
Moreover, we have demonstrated that
there exist eigenstates with vanishing energy, for $L\geq N(N-1)$. In addition,
we have considered 
the case of general attractive power-like two-particle potential
$r^k,\;k>-2$, and found ground state and its energy.

Note added:
After this work was completed a preprint by Kavoulakis et. al.\cite{22} 
appeared.
In the asymptotic limit $N\rightarrow\infty$ our eigenvalue $\Lambda_6^{(1)}$
from the relation (\ref{x}) is in full agreement with their findings 
in the presence of two octupole excitations.

Acknowledgment

This work was supported by the Ministry of Science and Technology of the
 Republic of Croatia under contract No. 00980103.

\end{document}